\documentclass[aps,twocolumn,floatfix]{revtex4}
\usepackage{amsfonts}
\usepackage{amsmath}
\usepackage{amssymb}
\usepackage{graphicx}
\begin{document}

\title{Thermally-assisted current-induced magnetization reversal in SrRuO$_{3}$}
\author{Yishai Shperber$^{1,*}$}
\author{Omer Sinwani$^{1,*}$}
\author{Netanel Naftalis$^{1}$}
\author{Daniel Bedau$^{2}$}
\author{James W. Reiner$^{2}$}
\author{Lior Klein$^{1}$}
\affiliation{$^{1}$Department of Physics, Nano-magnetism Research
Center, Institute of Nanotechnology and Advanced Materials,
Bar-Ilan
University, Ramat-Gan 52900, Israel}
\altaffiliation{Contributed equally to this work}
\affiliation{$^{2}$HGST, 3403 Yerba Buena Rd, San Jose, CA 95315 USA}

\begin{abstract}
We inject a sequence of 1 ms current pulses into uniformly magnetized patterns of the itinerant ferromagnet SrRuO$_{3}$ until a magnetization reversal is detected. We detect the effective temperature during the pulse and find that the cumulative pulse time $\tau$ required to induce magnetization reversal depends exponentially on $1/T$. In addition, we find that $\tau$ also depends exponentially on the current amplitude. These observations indicate current-induced magnetization reversal assisted by thermal fluctuations.

\end{abstract}

%\volumeyear{year}
%\volumenumber{number}
%\issuenumber{number}
%\eid{identifier}
%\date[Date text]{date}
%\received[Received text]{date}

%\revised[Revised text]{date}

%\accepted[Accepted text]{date}

%\published[Published text]{date}

%\pacs{73.50.Jt, 75.30.Gw}

\maketitle

\section{introduction}
The interaction of spin polarized current with magnetization gives rise to various fascinating spin-torque effects such as magnetization reversal of nanomagnets in junction configurations \cite{Berger1996,Slonczewski1996,Tsoi1998,Myers1999,Katine2000} and current-induced domain wall motion \cite{Berger1984,TataraDWM2004,Li2004,Yamaguchi2004,Feigenson2007,Boulle2008}. The interest in these effects is not only theoretically motivated, but also for their expected central role in novel spintronics devices, as they offer efficient and scalable methods to control  magnetic configurations on a nanometer scale.

An intriguing current-induced spin-torque effect is expected in a uniformly magnetized system with strong coupling between spin waves and current: in this case, currents above a threshold are expected to induce magnetization reversal \cite{Slonczewski1999,TataraNuc2005,Li2005,Korenblit2008,Togawa2008,Feigenson2008,Gerber2010}. The effect was observed for the itinerant ferromagnet SrRuO$_3$ (Curie temperature $T_c\sim 150 \ {\rm K}$) \cite{shperber2012}. However, while some features, such as a weak field dependence of the threshold current, were found consistent with existing models, the magnitude of the threshold current was an order of magnitude smaller than predicted.
Furthermore, the existing models ignore the role of thermal fluctuations which may be important for experiments performed at temperatures that are not much smaller than $T_c$.

Here, we explore the possible contribution of thermal fluctuations to current-induced magnetic instability in SrRuO$_3$ by studying the probability for reversal when the injected current is lower than the threshold current, while monitoring the effective temperature of the sample during the current injection. We determine the probability for reversal for a given current as a function of the temperature during the current injection, and the probability for reversal as a function of current for a given temperature. We find that for a given current the average cumulative reversal time ($\overline{\tau}$) depends exponentially on the inverse temperature, as expected from N$\rm{\acute{e}}$el Brown model \cite{Neel,Brown_pr},
%for thermally activated mechanism,
and for a given temperature $\overline{\tau}$ depends exponentially on the current. We find that the corresponding energy barrier is temperature dependent and that it is suppressed linearly with increasing current.

The results clearly show that for currents below the threshold current, current-induced magnetization reversal can be modeled by a thermally activated process with a current-dependent barrier.

\begin{figure}[t]
\includegraphics[scale=0.55, trim=50 200 50 250]{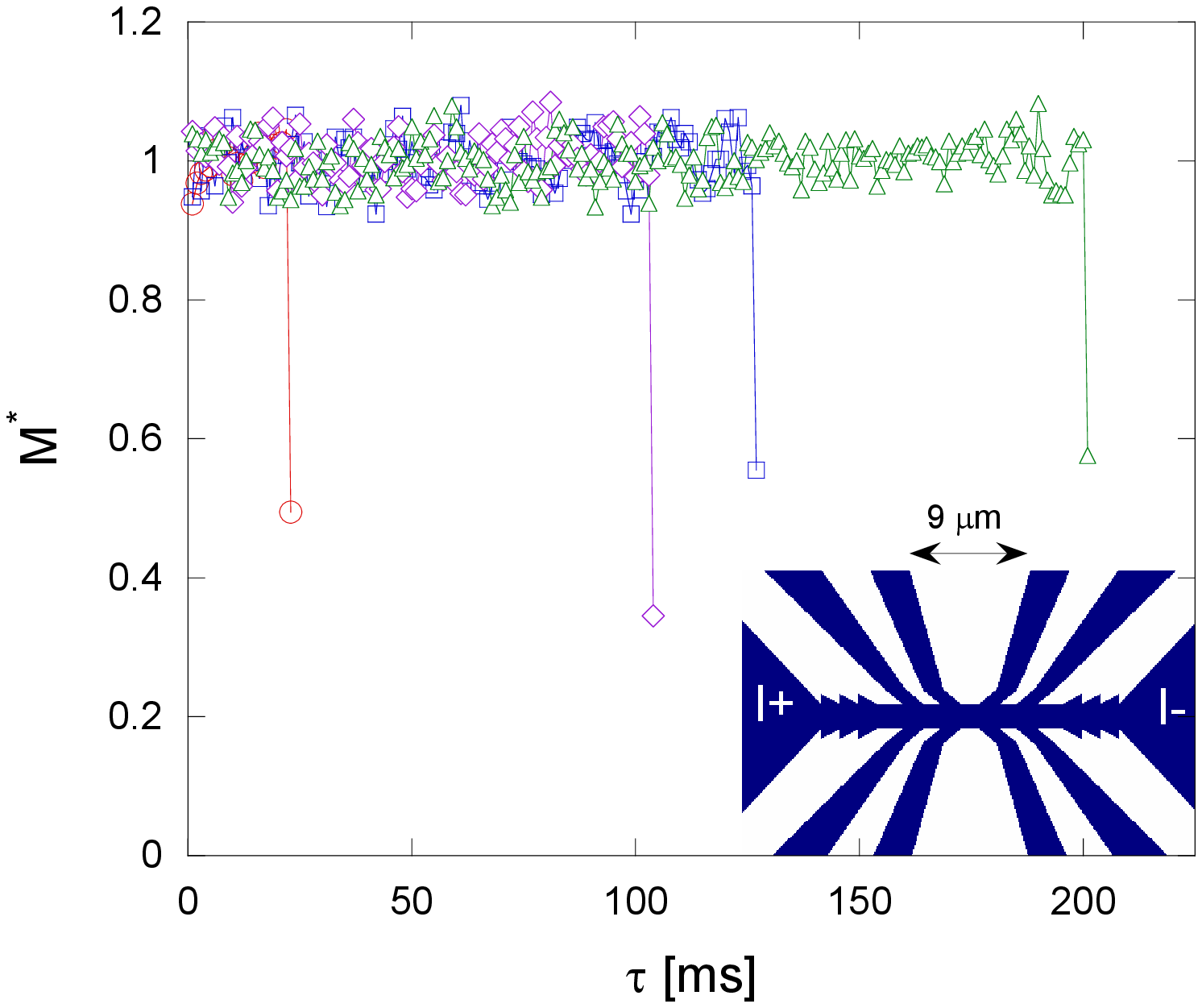}
\caption{The normalized average magnetization as measured in the last 50 $\mu$s of the 1 ms pulse vs. the accumulated pulse time $\tau$. The current amplitude is 4.28 mA, the external field is 500 Oe and the temperature is 107.8 K. The different symbols represent different sets of measurements. Inset: A sketch of the pattern.
}
\label{1}
\end{figure}

\section{experimental details}
Our samples are high-quality epitaxial films of SrRuO$_{3}$
 grown on slightly
miscut (2$^{\circ}$) SrTiO$_{3}$ substrates \cite{RevSrRuO3}.
The films are orthorhombic
($a=5.53$ \AA , $b=5.57$ \AA , $c=7.85$ \AA) with the
 $c$ axis in the film plane (perpendicular to the miscut direction) and the $a$ and $b$
axes are at 45$^{\circ}$ relative to the film plane.
The Curie temperature of the films is $\sim150$ K and they exhibit a
uniaxial magnetocrystalline anisotropy with the easy axis changing in the (001) plane between
the $b$ axis at $T\geq$$T_{c}$ to $30^\circ$ from the film normal at low temperatures \cite{Marshall,Klein1996}. The ratio between the resistivity at 300 K and the resistivity at the low temperature limit is greater than ten, indicative of the high quality of the films. The films are patterned for magnetotransport measurements, with a typical current path width of 1.5 $\mu$m, using e-beam lithography and $Ar^+$ ion milling. The data presented here are for a film thickness of 37.5 nm.

The current is injected in one millisecond pulses through the pattern as shown in the inset of Figure \ref{1}. The average temperature of the sample during the current pulse injection is determined with an accuracy of $\pm0.09$ K by measuring the longitudinal resistance during the current injection and using the known temperature dependence of the resistance of the sample (measured with a low current) \cite{shperber2012}. We cannot exclude a temperature variation along the current path on the order of 1 K. However, as the magnitude and form of the variation are expected to be practically the same in the temperature interval of our experiment, it may have a very minor effect on the analysis of our data.

The average magnetization state of the pattern is monitored by measuring
the transverse resistance which consists of both the ordinary Hall effect (OHE) and the anomalous Hall effect (AHE) related to the perpendicular component of the magnetic field and magnetization, respectively. Since the easy axis for magnetization in SrRuO$_3$ is tilted out of the film plane there is a perpendicular component of the magnetization in the remanent state when no field is applied \cite{Klein2000,Noam2011,Fang2003}.

\section{results and discussion}
Figure \ref{1} presents typical measurements of current induced magnetization reversal performed on a fully magnetized sample when the current amplitude is lower than the amplitude of the threshold current for the relevant temperature.
We inject a sequence of 1 ms pulses, separated by a 100 ms pause. During the last 50 $\mu$s of each pulse we measure the transverse resistance to detect changes in the average magnetization. If the magnetization has reversed, the sequence is stopped, the sample is fully magnetized again and the sequence of pulses is resumed. The time $\tau$ represents
the accumulated time of the current pulses and $M^*$ represents the normalized average magnetization extracted from the AHE signal.

The measurements presented here were performed for temperatures between 107 to 110 K with current amplitude between 4.15 to 4.31 mA (for our pattern, 1 mA corresponds to a current density of $\sim1.8\times10^{6}\frac{\rm{A}}{\rm{cm^2}}$). To facilitate the detection of a reversal event, we apply during the experiment a constant magnetic field of 500 Oe, opposite to the perpendicular component of the remanent magnetization, to assist propagation after nucleation. Since this field in the absence of current pulse induced nucleation only 25 K above the highest temperature at which our experiments are performed, one can conclude that this field has a negligible influence on the measured reversal events.

\begin{figure}[t]
\includegraphics[scale=0.45, trim=50 30 50 50]{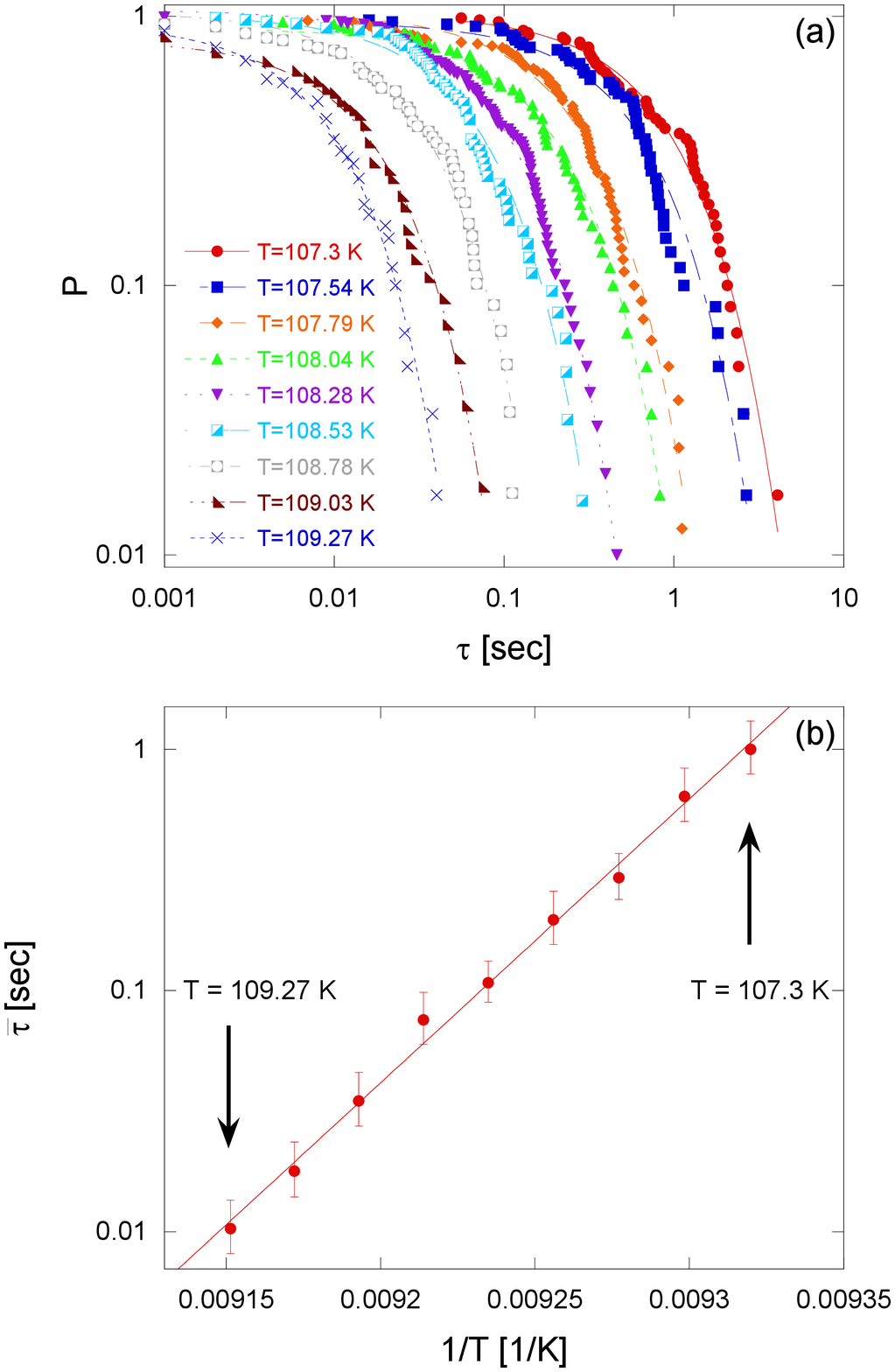}
\caption{(a) The probability P to remain fully magnetized vs. the accumulated time $\tau$, and (b) The average waiting time $\overline{\tau}$ for reversal as a function of the inverse temperature, for a current amplitude of 4.28 mA and an external field of 500 Oe opposite to the perpendicular component of the remanent magnetization. The error bars represent confidence interval of 95 $\%$.}
\label{2}
\end{figure}
Figure \ref{2} shows the results of experiments as shown in Figure \ref{1} with a current of
I=4.28 mA for temperatures between 107.3 to 109.3 K (the temperature is measured during the pulse, without current the temperature of the sample is $\sim$50 K lower). At each temperature the experiments were repeated 40-80 times.
Figure \mbox{\ref{2}(a)} shows the probability of the pattern to remain fully magnetized as a function of the accumulated pulse time. The lines represent the expected probability using the calculated probability for reversal after a single pulse and assuming an exponential distribution; namely, that the current pulses are uncorrelated events and the system has time to recover during the 100 ms between pulses. Based on the fits to exponential distributions, we extract $\overline{\tau}$, the average waiting time for reversal, for the given current amplitude and the given temperature.

Figure \ref{2}(b) shows the temperature dependence of \mbox{$\overline{\tau}$} for I=4.28 mA (symbols) as a function of $1/T$. The linear fit (with $\overline{\tau}$ in a logarithmic scale) indicates that

 \begin{equation}
\overline{\tau}=Ae^{E/(k_BT)},
\label{e0}
 \end{equation}
as expected for a thermally activated process.

\begin{figure}[t]
\includegraphics[scale=0.45, trim=50 150 50 250]{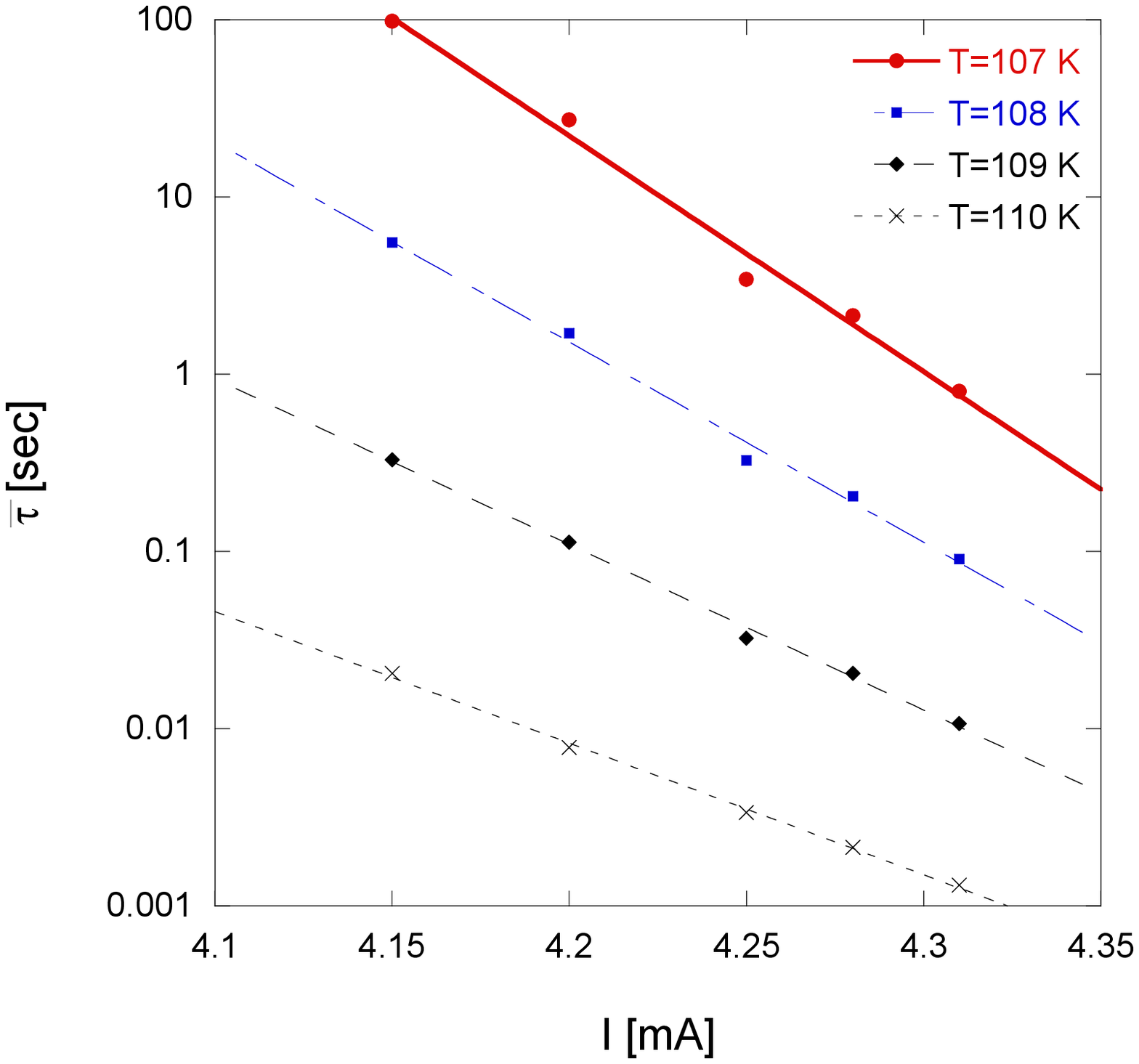}
\caption{The average waiting time for reversal as a function of the current amplitude for several temperatures. The
lines are fits to Eq. \ref{e1}.}
\label{3}
\end{figure}

Experiments as shown in Figure \ref{2} were repeated with multiple currents. Based on fits to Eq. \ref{e0} we extract by interpolation
$\overline{\tau}$ as a function of the current amplitude for given temperatures (see Figure \ref{3}). We note that a change of less than 5 percent in the current changes $\overline{\tau}$ by almost two orders of magnitude.

While a good fit with Eq. \ref{e0} is obtained for all currents, the fit parameters are current dependent. When the current increases, the parameter $E$ that represents the energy barrier for magnetization reversal decreases linearly, which suggests that
$E=E_b(1-I/I_c)$. In addition, we find that with increasing current the parameter $A$ increases exponentially. A possible source of this behavior is a temperature dependence of $E_b$ due to, for instance, the temperature dependence of the magnetization and the magnetic anisotropy in the temperature range of our experiments.

Since the temperature interval of our experiment is rather small (between 107 to 110 K), we use a linear approximation for the temperature dependence of $E_b$ which yields $E=E_b^0(1-\kappa \Delta T)(1-I/I_c)$, where $\Delta T=T-107$ K. Eq. \ref{e0} therefore takes the form:

\begin{equation}
\overline{\tau}=\tau_0e^{E_b^0(1-\kappa \Delta T)(1-I/I_c)/(k_BT)},
\label{e1}
 \end{equation}
where $\tau_0$ is a constant that is temperature and current independent, consistent with N$\rm{\acute{e}}$el Brown model. Interestingly, the dependence of the energy barrier $E_b$ on the current is very similar to the dependence observed in completely different cases, including the effect of the current on the energy barrier for depinning a domain wall \cite{Ravelosona2009} and the energy barrier for the switching of a nanomagnet in incorporated into a spin valve nanopillar \cite{Bedau2010}.

\begin{figure}[t]
\includegraphics[scale=0.45, trim=50 150 50 250]{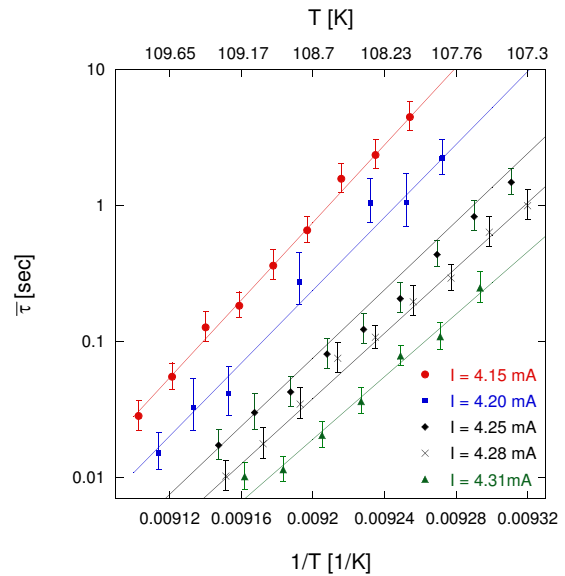}
\caption{The average waiting time for reversal for several current amplitudes as a function of the inverse temperature (lower axis) - the corresponds temperature is also presented (upper axis). The lines are fits to Eq. \ref{e1}. The error bars represent confidence interval of 95 $\%$.}
\label{4}
\end{figure}
 Using Eq. \ref{e1} we fit the data obtained for all the currents with the same fitting parameters (see Figure \ref{4}).
 We obtain the best fit with $\tau_0=4.8\times10^{-9}$ s, $E_b^0= 1.32$ eV, $\kappa=0.11\rm{\frac{1}{K}}$ and $I_{c}=4.97$ mA, in agreement with the measured threshold current \cite{shperber2012}.
Using a bootstrap method, we find that the error bars for a confidence interval of 95 percent are $E_b^0\sim1.28-1.41$ eV, $\kappa\sim0.07-0.15\rm{\frac{1}{K}}$ and $I_{c}\sim4.7-5.4$ mA. As small variations in these parameters change $\tau_0$ exponentially, its error bars are significantly larger. We find that for a confidence interval of 50$\%$ the possible range of $\tau_0$ is~$\sim1\times10^{-10}-8\times10^{-8}$ s.

To estimate the nucleation volume, we use the relation $E_b^0=K_uV$ expected for coherent rotation, where $K_u$ is the magnetic anisotropy energy density related to the uniaxial magnetocrystalline anisotropy. At low temperatures $K_u \sim 5.6\times10^{-3}\rm{\frac{eV}{nm^3}}$ (extracted from the measured anisotropy field of $\sim$7 T \cite{Kerr} and saturation magnetization  $M_{s}\sim250$ kA/m \cite{RevSrRuO3}) and using this value yields $V \sim 240$ nm$^3$. Assuming a nucleation volume of cylindrical shape with the film thickness (37.5 nm) as the cylinder height, the estimated radius of the cylinder is $\sim$ 1.5 nm, on the order of the  domain wall width.
We should, however, take into consideration that $E_b^0$ is the barrier at T $=$ 107 K, where the magnetization is $\sim$ 0.7 of its saturation value and  $K_u$ may also be suppressed
\cite{anisotropy1,anisotropy2}. Therefore, the volume could be larger although still on the same order.

\section{Conclusions}
In conclusion, we have studied the current-induced switching in a uniformly magnetized
ferromagnetic \mbox{SrRuO$_3$} sample close to the Curie temperature, and we have
demonstrated that the switching process can be described as a constant
prefactor N$\rm{\acute{e}}$el Brown model with first order corrections to the energy
barrier for current and temperature.
%We demonstrate that the temperature dependence of the switching
%energy barrier has linear dependence on temperature as well as current.
The observed dependence of the energy barrier on current makes it possible
to modulate the energy barrier in itinerant ferromagnets, which might
offer opportunities for further study of the switching process and for the development of novel spintronics devices.

\section{acknowledgments}
L.K. acknowledges support by the Israel Science Foundation founded by the Israel Academy of Sciences and Humanities. J.W.R. grew
the samples at Stanford University in the laboratory of M. R. Beasley.


\begin{thebibliography}{9}
\bibitem {Berger1996}L. Berger, Phys. Rev. B \textbf{54}, 9353 (1996).
\bibitem {Slonczewski1996}J. C. Slonczewski, J. Magn. Magn. Mater. \textbf{159}, L1 (1996).
\bibitem {Tsoi1998}M. Tsoi, A. G. M. Jansen, J. Bass, W.-C. Chiang, M. Seck, V. Tsoi, and P.
Wyder, Phys. Rev. Lett. \textbf{80}, 4281 (1998).
\bibitem {Myers1999}E. B. Myers, D. C. Ralph, J. A. Katine, R. N. Louie, and
R. A. Buhrman, Science \textbf{285}, 867 (1999).
\bibitem {Katine2000}J. A. Katine, F. J. Albert, R. A. Buhrman, E. B. Myers, and D. C. Ralph,
Phys. Rev. Lett. \textbf{84}, 3149 (2000).
\bibitem {Berger1984}L. Berger, J. Appl. Phys. \textbf{55}, 1954 (1984).
\bibitem {TataraDWM2004} G. Tatara, and H. Kohno, Phys. Rev. Lett. \textbf{92}, 086601 (2004).
\bibitem {Li2004} S. Zhang, and Z. Li, Phys. Rev. Lett. \textbf{93}, 127204 (2004).
\bibitem {Yamaguchi2004}A. Yamaguchi, T. Ono, S. Nasu, K. Miyake, K. Mibu, and T. Shinjo, Phys. Rev. Lett. \textbf{92}, 077205 (2004).
\bibitem {Feigenson2007}M. Feigenson, J. W. Reiner, and L. Klein, Phys. Rev. Lett. \textbf{98}, 247204 (2007).
\bibitem {Boulle2008} O. Boulle, J. Kimling, P. Warnicke, M. Kl\"{a}ui, U. R\"{u}diger, G. Malinowski, H. J. M. Swagten, B. Koopmans, C. Ulysse and G. Faini, Phys. Rev. Lett. \textbf{101}, 216601 (2008).
\bibitem {Slonczewski1999}J. C. Slonczewski, J. Magn. Magn. Mater. \textbf{195}, L261 (1999).
\bibitem {Li2005}Z. Li, J. He, and S. Zhang, J. Appl. Phys. \textbf{97}, 10C703 (2005).
\bibitem {TataraNuc2005}J. Shibata, G. Tatara, and H. Kohno, Phys. Rev. Lett. \textbf{94}, 076601 (2005).
\bibitem {Korenblit2008}I. Ya. Korenblit, Phys. Rev. B \textbf{77}, 100404(R) (2008).
\bibitem {Togawa2008}Y. Togawa, T. Kimura, K. Harada, T. Matsuda, A. Tonomura, T. Akashic, and Y. Otanib, Appl. Phys. Lett. \textbf{92}, 012505 (2008).
\bibitem {Feigenson2008}M. Feigenson, J. W. Reiner, and L. Klein, J. Appl. Phys. \textbf{103}, 07E741 (2008).
\bibitem {Gerber2010} O. Riss, A. Gerber, I. Ya. Korenblit, M. Karpovsky, S. Hacohen-Gourgy, A. Tsukernik, J. Tuaillon-Combes, P. M\'{e}linon, and A. Perez, Phys. Rev. B \textbf{82}, 144417 (2010).
\bibitem{shperber2012} Y. Shperber, D. Bedau, J. W. Reiner, and L. Klein, Phys. Rev. B \textbf{86}, 085102 (2012).
\bibitem{Neel} M. L. N\'{e}el, Ann. Geophys. \textbf{5}, 99 (1949).
\bibitem{Brown_pr} W. F. Brown, Phys. Rev. \textbf{130}, 1677 (1963).
\bibitem {RevSrRuO3} G. Koster, L. Klein, W. Siemons, G. Rijnders, J. S. Dodge, C. B. Eom, D. H. A. Blank, and Malcolm R. Beasley, Rev. Mod. Phys. \textbf{84}, 253 (2012).
\bibitem {Marshall}A. F. Marshall, L. Klein, J. S. Dodge, C. H. Ahn, J. W. Reiner, L. Mieville, L. Antognazza, A. Kapitulnik, T. H. Geballe, and M. R. Beasley, J. Appl. Phys. \textbf{85},
4131 (1999).
\bibitem{Klein1996}L. Klein, J. S. Dodge, C. H. Ahn, J. W. Reiner, L. Mieville, T. H. Geballe, M. R. Beasley, and A. Kapitulnik, J. Phys.: Condens. Matter \textbf{8}, 10111 (1996).
\bibitem {Klein2000}L. Klein, Y. Kats, A. F. Marshall, J.W. Reiner, T. H. Geballe, M. R. Beasley, and A. Kapitulnik, Phys. Rev. Lett. \textbf{84}, 6090 (2000).
\bibitem {Noam2011}N. Haham, Y. Shperber, M. Schultz, N. Naftalis, E. Shimshoni, J. W. Reiner, and L. Klein, Phys. Rev. B \textbf{84}, 174439 (2011).
\bibitem {Fang2003} Z. Fang, N. Nagaosa, K. S. Takahashi, A. Asamitsu, R. Mathieu, T. Ogasawara, H. Yamada, M. Kawasaki, Y. Tokura, and K. Terakura, Science \textbf{302}, 92 (2003).
\bibitem{Ravelosona2009}J. Cucchiara, Y. Henry, D. Ravelosona, D. Lacour, E. E. Fullerton, J. A. Katine, and S. Mangin, Appl. Phys. Lett. \textbf{94}, 102503 (2009).
\bibitem{Bedau2010}D. Bedau, H. Liu, J. Z. Sun, J. A. Katine, E. E. Fullerton, S. Mangin, and A. D. Kent, Appl. Phys. Lett. \textbf{97}, 262502 (2010).
%\bibitem{Omer2012}O. Sinwani, J. W. Reiner, L. Klein, Phys. Rev. B \textbf{86}, 100403(R) (2012).
\bibitem{anisotropy1}O. N. Mryasov, U. Nowak, K. Y. Guslienko, and R. W. Chantrell, Europhys. Lett., \textbf{69} (5), pp. 805–811 (2005).
\bibitem{anisotropy2}J. B. Staunton, L. Szunyogh, A. Buruzs, B. L. Gyorffy, S. Ostanin, and L. Udvardi, Phys. Rev. B \textbf{74}, 144411 (2006).
\bibitem{Kerr} M. C. Langner, C. L. S. Kantner, Y. H. Chu, L. M. Martin, P. Yu, J. Seidel, R. Ramesh and J. Orenstein, Phys. Rev. Lett. \textbf{102}, 177601 (2009).

%\bibitem {Li2006} Z. Li, H. He, and S. Zhang, J. Appl. Phys. \textbf{99}, 08Q702 (2006).

%\bibitem{Cullity} B. D. Cullity, C. D. Graham, Introduction to Magnetic Materials, Second Edition, John Wiley and Sons, p.384 (2009).

%%%%%\bibitem {Miron}Miron $et al$, Nature \textbf{476}, 7359, 189-U88 (2011).
\end{thebibliography}
\end{document}